\documentstyle[sprocl]{article}
\def\be{\begin{eqnarray}}
\def\en{\end{eqnarray}}
\def\non{\nonumber}
\def\la{\langle}
\def\cs{{\Lambda_c\Lambda}}
\def\bc{{\Lambda_b\Lambda_c}}
\def\ra{\rangle}

\def\lsim{ {\ \lower-1.2pt\vbox{\hbox{\rlap{$<$}\lower5pt\vbox{\hbox{$\sim$}
}}}\ } }
\def\gsim{ {\ \lower-1.2pt\vbox{\hbox{\rlap{$>$}\lower5pt\vbox{\hbox{$\sim$}
}}}\ } }

\def\pr{{\sl Phys. Rev.}~}
\def\prl{{\sl Phys. Rev. Lett.}~}
\def\pl{{\sl Phys. Lett.}~}

\begin{document}
\title{BARYONIC FORM FACTORS IN THE NONRELATIVISTIC QUARK MODEL}
\author{ HAI-YANG CHENG }
\address{Institute of Physics, Academia Sinica, Taipei, Taiwan 115, Republic
of China}
\maketitle\abstracts{
    Weak current-induced baryonic form factors at zero recoil are evaluated
in the rest frame of the heavy parent baryon using the nonrelativistic quark
model. Contrary to previous similar work in the literature, our quark model
results do satisfy the constraints imposed by heavy quark symmetry
for heavy-heavy baryon transitions.
Assuming a dipole $q^2$ behavior, we have applied the quark
model form factors to nonleptonic, semileptonic
and weak radiative decays of the heavy baryons.  }

     The general expression for the ${1\over 2}^+\to {1\over 2}^+$ baryonic 
transition $B_i\to B_f$ reads
\be
\la B_f(p_f)|V_\mu-A_\mu|B_i(p_i)\ra  & = & \bar{u}_f
[f_1(q^2)\gamma_\mu+if_2(q^2)\sigma_{\mu\nu}q^\nu+f_3(q^2)q_\mu    \\
&&{}-(g_1(q^2)\gamma_\mu+ig_2(q^2)\sigma_{\mu\nu}q^\nu+g_3(q^2)q_\mu)\gamma_5]
u_i,  \non
\en
where $q=p_i-p_f$. In the heavy-quark limit, the form factors $f_i$ and $g_i$
are related to three baryonic Isgur-Wise functions: $\zeta(\omega)$ for 
antitriplet-antitriplet transition,
and $\xi_1(\omega),~\xi_2(\omega)$ for sextet-sextet transition. 
In general, it is difficult to estimate the $1/m_Q$ corrections to hadronic
form factors in the heavy quark effective theory (HQET).
However, a tremendous simplification occurs in the
antitriplet-antitriplet heavy baryon transition, e.g., $\Lambda_b\to
\Lambda_c$: $1/m_Q$ corrections only amount to renormalizing the function
$\zeta(\omega)$ and no further new function is needed \cite{Georgi}. This
simplification stems from the fact that the chromo-magnetic operator does not
contribute to $\Lambda_b\to \Lambda_c$ and that the diquark of the antitriplet
heavy baryon is a spin singlet. 

   Going beyond the antitriplet-antitriplet heavy baryon transition, the
predictive power of HQET for form factors at order $1/m_Q$ is lost owing to
the fact that $1/m_Q$ corrections due to wave function modifications arising
from $O_1$ and especially $O_2$ are not calculable by perturbative QCD.
Therefore, it is appealing to have model calculations which enable us to
estimate the $1/m_Q$ corrections for other baryon form factors.
To our knowledge, two different quark-model calculations \cite{Perez,Sing} 
are available in the literature.
Unfortunately, none of the calculations presented
in \cite{Perez,Sing} is in agreement with the predictions of HQET. For example,
several heavy quark symmetry relations between baryon form factors are
not obeyed in \cite{Perez}. While this discrepancy is
resolved in \cite{Sing}, its prediction for $\Lambda
_b\to\Lambda_c$ (or $\Xi_b\to\Xi_c$) form factors at order $1/m_Q$ is still
too large by a factor of 2 when compared with HQET.

   In \cite{CT} we have shown that our
prescription of quark model calculations does incorporate the features
of heavy quark symmetry and hence can be applied to compute
baryon form factors beyond the arena of HQET. For example, for $\Lambda_b
\to\Lambda_c$ transition we obtain
\be
f_1^\bc(q^2_m)=g_1^\bc(q^2_m) &=& 1+{\Delta m\bar{\Lambda}\over 4}\left({1
\over m_{\Lambda_b}m_c}-{1\over m_{\Lambda_c}m_b}\right),   \non \\
f_2^\bc(q^2_m)=g_3^\bc(q^2_m) &=& -{\bar{\Lambda}\over 4}\left({1\over m_{
\Lambda_b}m_c}+{1\over m_{\Lambda_c}m_b}\right),    \\
f_3^\bc(q^2_m)=g_2^\bc(q^2_m) &=& -{\bar{\Lambda}\over 4}\left({1\over m_{
\Lambda_b}m_c}-{1\over m_{\Lambda_c}m_b}\right),   \non
\en
at zero reoil, where $q_m=(m_{\Lambda_b}-m_{\Lambda_c})^2$, and $\bar{\Lambda}
=m_{\Lambda_c}-m_c$. The above results agree with the HQET predictions up to
the zeroth order of $\alpha_s$.

   Two remarks are in order. First, the baryonic form factor is proportional
to a flavor factor $N_{fi}=_{\rm flavor}\!\!\!\!\la B_f|b_q^\dagger b_Q|B_i\ra
_{\rm flavor}$, which is equal to unity for heavy-to-heavy transition, but
less than unity for heavy-to-light transition. For example,
$N_{\Lambda_c\Lambda}=1/\sqrt{3}$.
In the literature it is customary to replace the $s$ quark
in the baryon $\Lambda$ by the heavy quark $Q$ to obtain the wave function
of the $\Lambda_Q$. However, this amounts to assuming SU(4) or SU(5) flavor
symmetry. Since SU(N)-flavor symmetry with $N>3$
is badly broken, the flavor factor $N_{\Lambda_Q\Lambda}$ is no longer unity.
Indeed, if $N_\cs$ were equal to
one, the predicted rate for $\Lambda_c\to\Lambda e^+\nu_e$ would have been
too large by at least a factor of 2 ! Second, as
the conventional practice, we make the pole dominance assumption for
the $q^2$ dependence to extrapolate the form factors from maximum $q^2$ to
the desired $q^2$ point. We argued that a dipole $q^2$ behavior is more
preferred since it is close to the baryonic Isgur-Wise function calculated
recently. Nevertheless, one should bear in mind that the assumption of
pole dominance for form factors is probably too simplified and this problem
remains unresolved.

  We have applied the quark model form factors to nonleptonic, semileptonic
and weak radiative decays of the heavy baryons \cite{CT}. In the heavy
$c$-quark limit, there are two independent form factors in $\Lambda_c
\to\Lambda$ transition 
\be
\la\Lambda(p)|\bar{s}\gamma_\mu(1-\gamma_5)c|\Lambda_c(v)\ra=\,\bar{u}
_{_\Lambda}\left(F_1^{\Lambda_c\Lambda}(v\cdot p)+v\!\!\!/ F_2^{\Lambda_c
\Lambda}(v\cdot p)\right)\gamma_\mu(1-\gamma_5)u_{_{\Lambda_c}}.
\en
We found that
\be
R\equiv F_1^{\Lambda_c\Lambda}/F_2^{\Lambda_c\Lambda}=-1/\left(1+4{m_s\over
\bar{\Lambda}}\right)=-0.23\,,
\en
with $\bar{\Lambda}=m_{\Lambda_c}-m_s$,
in accord with the CLEO result \cite{CLEO}
$R=-0.25\pm 0.14\pm 0.08$. For the semileptonic
decay $\Lambda_c\to\Lambda e^+\nu_e$, the decay asymmetry parameter
is obtained to be $\la\alpha\ra=-0.84$, in good agreement with experiment
\cite{PDG} $\la\alpha\ra =-0.82^{+0.11}_{-0.07}$.
The decay rate is predicted to be
\be
\Gamma(\Lambda_c\to\Lambda e^+\nu_e)=\,(N_\cs)^2\times 2.11\times 10^{11}s^
{-1}=\,7.1\times 10^{10}s^{-1},
\en
while experimentally \cite{PDG} $\Gamma(\Lambda_c\to\Lambda e^+\nu_e)_{\rm 
expt}=\,(
11.2\pm 2.4)\times 10^{10}s^{-1}$. The presence of the flavor suppression 
factor $N_{\Lambda_c\Lambda}$, which is missed in most literature,
will of course affect the predictions on the decay rates of
many decay modes involving a transition from heavy to light baryons.

   Generally it is very difficult to tackle the nonleptonic weak decays of
the baryons because of the presence of nonspectator $W$ exchange effects
manifested as pole contributions. Nevertheless, there are two 
decay modes of great interest, namely $\Lambda_c\to p\phi$ and 
$\Lambda_b\to\Lambda J/\psi$
which are theoretically very clean in the sense that they proceed only through
the internal $W$-emission diagram. Noting $N_{\Lambda_c p}={1\over\sqrt{2}}$
and $N_{\Lambda_b\Lambda}={1\over\sqrt{3}}$ and using the factorization
approach, we found $Br(\Lambda_c\to p\phi)=7.1\times 10^{-4}$, which agrees
with the experimental result \cite{PDG} $(1.06\pm 0.33)\times 10^{-3}$.
The branching ratio of $\Lambda_b\to\Lambda J/\psi$ is predicted to be 
$2.1\times 10^{-4}$, which is two orders
of magnitude smaller than the UA1 observation \cite{UA1} but consistent with 
the most recent CDF measurement  \cite{CDF} $Br(\Lambda_b\to\Lambda J/\psi)=
(4.2\pm 1.8\pm 0.7)\times 10^{-4}$.

\section*{Acknowledgments}
   This work was done in collaboration with B. Tseng and supported in part 
by the National Science Council of ROC
under Contract No. NSC85-2112-M-001-010.

\newcommand{\bi}{\bibitem}

\end{document}